\pdfoutput=1
\documentclass[sigconf]{acmart}
\acmDOI{x}
\usepackage{booktabs} % For formal tables
\usepackage{graphicx}
\usepackage{ifthen}

\usepackage[utf8]{inputenc} % allow utf-8 input
\usepackage[T1]{fontenc}    % use 8-bit T1 fonts
\usepackage{url}            % simple URL typesetting
\usepackage{booktabs}       % professional-quality tables
\usepackage{amsfonts}       % blackboard math symbols
\usepackage{dsfont}

\usepackage{nicefrac}       % compact symbols for 1/2, etc.
\usepackage{microtype}      % microtypography
\usepackage{amsmath}
\usepackage{bm}
\usepackage{subcaption}
\usepackage{mathrsfs}
\usepackage{xcolor}
\usepackage{enumitem}
\usepackage{algorithm}
\usepackage{algorithmic}
\usepackage{multirow}
\usepackage{hyperref}
\hypersetup{colorlinks,allcolors=black}
\usepackage{refcount}

\usepackage{balance}

\DeclareMathOperator*{\argmin}{arg\,min}

\begin{document}
\title{Density-Ratio Based Personalised Ranking from~Implicit~Feedback}

\author{Riku Togashi}
\affiliation{
  \institution{CyberAgent, Inc.}
  \city{Tokyo} 
  \country{Japan}
}
\email{rtogashi@acm.org}

\author{Masahiro Kato}
\affiliation{
  \institution{CyberAgent, Inc.}
  \city{Tokyo} 
  \country{Japan}
}
\email{kato_masahiro@cyberagent.co.jp}

\author{Mayu Otani}
\affiliation{
  \institution{CyberAgent, Inc.}
  \city{Tokyo} 
  \country{Japan}
}
\email{otani_mayu@cyberagent.co.jp}

\author{Shin'ichi Satoh}
\affiliation{
  \institution{CyberAgent, Inc.}
  \city{Tokyo} 
  \country{Japan}
}
\email{satoh@nii.ac.jp}

\renewcommand{\shortauthors}{
}

\begin{abstract}
Learning from implicit user feedback is challenging as we can only observe positive samples but never access negative ones.
Most conventional methods cope with this issue by adopting a pairwise ranking approach with negative sampling.
However, the pairwise ranking approach has a severe disadvantage in the convergence time owing to the quadratically increasing computational cost with respect to the sample size;
it is problematic, particularly for large-scale datasets and complex models such as neural networks.
By contrast, a pointwise approach does not directly solve a ranking problem, and is therefore inferior to a pairwise counterpart in top-$K$ ranking tasks;
however, it is generally advantageous in regards to the convergence time.
This study aims to establish an approach to learn personalised ranking from implicit feedback, which reconciles the training efficiency of the pointwise approach and ranking effectiveness of the pairwise counterpart.
The key idea is to estimate the ranking of items in a pointwise manner;
we first reformulate the conventional pointwise approach based on density ratio estimation and then incorporate the essence of ranking-oriented approaches (e.g. the pairwise approach) into our formulation.
Through experiments on three real-world datasets, we demonstrate that our approach not only dramatically
reduces the convergence time (one to two orders of magnitude faster) but also significantly improving the ranking performance.
\end{abstract}




\maketitle

\section{Introduction}
Recommender systems play a vital role in alleviating information-overload problems by delivering the relevant items buried in an enormous amount of non-relevant ones.
For tracking up-to-date interests of users in modern web applications such as E-commerce and social media (e.g. SNS),
implicit user feedback (e.g. views, clicks, and purchases) is a key source.
Implicit feedback is automatically collected as users' behavioural traces, reflecting current interests of users.
However, there are several difficulties owing to the \emph{one-class problem}~\cite{pan2008one};
that is, we can observe only positive feedback.
For instance, we can access a click log of a user (i.e. positive feedback), but we can never observe negative feedback from a user,
which indicates that the user dislikes an item and will not click it even in the future.

For handling such positive-only data, conventional methods
mainly adopt the pairwise approach with negative sampling~\cite{rendle2009bpr},
where pseudo-negative samples are selected from unobserved ones.
However, it is well-known that the pairwise approach has a disadvantage in convergence time.
This is because that the loss function is computed over all possible pairs of items that grow quadratically in the number of items, and a large proportion of pairs are not informative for the model in later iterations~\cite{zhang2013optimizing,rendle2014improving}.
The inefficiency of the pairwise approach in model training is rather severe for large-scale datasets.

In contrast to such sampling-based approaches, whole-data-based methods~\cite{devooght2015dynamic,liang2016modeling} have been developed on the basis of weighted matrix factorisation (WMF)~\cite{pan2008one,hu2008collaborative}.
The methods leverage all the unobserved samples as weak negative instances by weighting instead of sampling~\cite{liang2016modeling}.
Recent methods based on the non-sampling approach have demonstrated efficient model training based on the loss function of the WMF, and achieved excellent convergence time even with complex models based on neural networks~\cite{chen2019social,chen2019efficient,chen2020efficient,chen2020jointly}.
However, from the top-$K$ ranking perspective, 
they are often inferior to pairwise counterparts because their loss functions are based on pointwise regression, which is inappropriate for a ranking problem~\cite{cremonesi2010performance}.
In addition, they require to leverage all the users or items in each training step, and thus, can be inefficient or even infeasible for large-scale data.

Taking these shortcomings of the conventional methods into account,
we assume that there are three requirements for a method without compromise in both training efficiency and ranking effectiveness.
First, a method should optimise a criterion that aligns with top-$K$ ranking measures while overcoming the inefficiency issue of the pairwise approach.
Second, the model training must be conducted through mini-batch stochastic gradient descent (SGD) on arbitrarily sub-sampled users/items,
which is a standard method for training neural-network-based models on large-scale data.
Finally, all of these must be feasible on one-class data, namely, implicit user feedback.

In this study, we take a direct approach to realise a method that satisfies the above requirements.
The crux of our approach is to estimate the order of items in a pointwise manner.
To this end, we first interpret the conventional whole-data-based methods from the viewpoint of semi-supervised learning.
Then, we propose a risk based on semi-supervised density ratio estimation (DRE),
which is a more appropriate task to be solved for top-$K$ recommendation than class probability estimation (i.e. classification).
Considering the intrinsic behaviour of evaluation measures in top-$K$ ranking,
we also introduce a sample weighting strategy for inducing a top-weighted property in the risk.
We then derive an empirical risk approximation to enable SGD-based model training on arbitrarily sub-sampled users/items.
Through experiments on three real-world datasets, we demonstrate that the method based on our framework can dramatically reduce the convergence time while achieving comparable or even substantially higher ranking performance than conventional methods.

To summarise, our main contributions are as follows:
\begin{itemize}
    \item We develop a DRE-based framework for personalised ranking, which is designed to achieve efficient training of complex models on large-scale data. Our framework incorporates a ranking-aware sample weighting strategy into a pointwise approach.
    \item We propose a DRE-based risk as a generalised form and thereby enabling flexible extension for further advanced methods. We also provide an implementation of our framework built upon a state-of-the-art model (i.e. LightGCN~\cite{he2020lightgcn}), which we call \emph{DRE-based Graph Neural Collaborative Filtering (DREGN-CF)}.
    \item We demonstrate the superiority of the DREGN-CF in terms of both ranking effectiveness and training efficiency through experiments on real-world datasets. We also discuss the characteristics of our DRE-based approach through extensive ablation analyses.
\end{itemize}

\section{Related Work}
\subsection{Personalised Recommendation with Implicit User Feedback}
Most conventional methods for personalised recommendation can be classified into two fundamental approaches;
(1) pointwise approach~\cite{pan2008one,hu2008collaborative,koren2009matrix}; and (2) pairwise approach~\cite{rendle2009bpr}.
The pointwise approach typically solves a binary classification problem to predict if a user will prefer an item or not.
The pairwise approach solves another binary classification problem to predict which of the given two items a user will prefer.

Methods based on the pairwise approach for implicit feedback rely on negative sampling strategies that exploit unobserved samples as pseudo-negatives.
The choice of a negative sampling strategy is quite important for achieving optimal ranking performance and efficient training~\cite{zhang2013optimizing,rendle2014improving}.
Early NS strategies select negatives according to a fixed distribution~\cite{rendle2009bpr}.
For further improving training efficiency, adaptive sampling strategies are also proposed later~\cite{zhang2013optimizing,rendle2014improving}.
Those methods adaptively mine hard negative samples by regarding the status of a model.
Advanced methods that utilise an adversarial sampler model (a.k.a. the generator)~\cite{IRGAN,AdvIR} select informative negatives that distribute around the decision boundary of a ranker model;
this concept is incorporated into our sample weighing strategy.

Methods without negative sampling have also been established.
WMF is a fundamental model that leverages the whole data with sample weighting~\cite{pan2008one,hu2008collaborative}.
Advanced methods based on matrix factorisation have been developed~\cite{pan2009mind,devooght2015dynamic,liang2016modeling,saito2020unbiased}.
As such whole-data-based methods typically suffer from computational costs due to the use of full user-item interaction matrix, various methods have been examined for efficiently handling the huge matrix by assuming uniform weights for the unobserved samples~\cite{hu2008collaborative,devooght2015dynamic}.
The conventional methods are mainly based on alternating least squares (ALS)~\cite{hu2008collaborative,pan2008one,devooght2015dynamic,he2019fast}.
On the other hand, a weighting strategy for unobserved samples in a whole-data-based method plays a similar role to the negative sampling strategy in a sampling-based counterpart.
Several methods with non-uniform weighting have also been proposed whereas they experience severe computational complexity owing to the dense weight matrix for all possible user-item pairs~\cite{pan2008one,pan2009mind}.
He et al.~\cite{he2019fast} recently proposed a fast ALS-based method for non-uniform weights for unobserved samples.
Chen et al. recently proposed 
another direction of WFM-based methods, namely, the non-sampling approach, which optimises the loss function of the WMF with mini-batch SGD~\cite{chen2019efficient,chen2020efficient,chen2020jointly}.
However, two challenges remain in the non-sampling approach;
(1) its pointwise regression loss can impair top-$K$ ranking performance~\cite{cremonesi2010performance}; and
(2) it leverages all the users or items in each training step and can be infeasible for training complex models (e.g. graph neural networks (GNNs)~\cite{NGCF}) on large-scale data, particularly in terms of memory space.

The listwise approach is another direction of ranking-oriented methods~\cite{shi2010list,huang2015listwise}.
In particular, methods based on variational autoencoders (VAEs) can be considered as one of the mainstream listwise approach based on neural networks~\cite{liang2018variational,steck2019embarrassingly,shenbin2020recvae}.
Liang et al.~\cite{liang2018variational} proposed a method for modelling a user's implicit feedback by a multinomial distribution across all items for the user; therefore, the model restricts the sum of interaction probabilities for a user to one, thereby representing \emph{relative} interaction probabilities of items for a user.
This approach can be regarded as the extension to top-$K$ ranking optimisation of the classical listwise methods that typically maximise top-$1$ recall, such as ListNet~\cite{cao2007learning}.
However, a way to combine those VAE-based methods and an arbitrary model architecture (e.g. GNNs) is not obvious as they are based on a certain model architecture.
Moreover, those methods also incur severe computational costs when the number of items is large.

In this paper, we propose a sampling-based and pointwise approach.
In order to alleviate the scalability issue of the non-sampling and listwise approaches,
we first sub-sample both users and items with a simple and static sampling strategy, and then utilise all possible user-item pairs in a mini-batch including unlabelled samples.
To enjoy the effectiveness of sampling strategies with the adversarial sampler model for the pairwise approach and adapt this for our pointwise approach, 
we introduce a non-uniform ranking-aware weighting strategy for both observed and unobserved samples.
Furthermore, we propose a fundamentally different loss function based on semi-supervised DRE to tackle the ineffectiveness of the pointwise approach in terms of top-$K$ ranking performance.

\subsection{Learning to Rank and Density Ratio Estimation}
Many studies have proposed methods for improving top-$K$ ranking performance by introducing learning-to-rank (LTR) approaches.
Christakopoulou et al.~\cite{christakopoulou2015collaborative} extensively examined top-emphasised pairwise loss functions~\cite{rudin2009p,agarwal2011infinite,rakotomamonjy2012sparse} for collaborative filtering.
Although they primarily investigated the family of pairwise losses, 
the proof of the equivalence of a top-emphasised pairwise loss function~\cite{rudin2009p} and classification counterpart (i.e. $p$-classification loss) has been provided by Ertekin et al.~\cite{ertekin2011equivalence}.
From a different perspective, Menon et al.~\cite{menon2016linking} suggested that LTR is achievable through solving DRE and demonstrated the top-weighted property of DRE-based loss functions;
in their experiments, a method based on a DRE approach, namely, the unconstrained least squared importance fitting (uLSIF)~\cite{kanamori2009least}, outperforms the one based on the $p$-classification loss in terms of ranking measures such as average precision.
In this paper, we propose a DRE-based LTR approach for taking the advantage of DRE in top-$K$ ranking, as suggested by Menon et al.~\cite{menon2016linking}.
The highlight of our approach in the context of DRE is that
we specialise a conventional DRE risk for personalised ranking from implicit feedback with sample- and class-dependent weights and user-wise optimisation strategy.
Furthermore, we propose a ranking-aware sample weighting strategy for explicitly inducing a top-weighted property in our proposed risk.
Considering settings with large-scale data and complex models, 
we derive a loss function through empirical risk approximation for enabling SGD-based optimisation.

Several works have explored LTR applications of DRE.
Acharyya et al.~\cite{acharyya12} proposed an LTR method for explicit settings that formulates the ranking problem as the minimisation of the Bregman divergence, which is a wide class of divergences (e.g., Kullback-Leibler divergence~\cite{kullback1951information}) and closely related to DRE~\cite{sugiyama2012density}.
Providing a theoretical background to the work of Acharyya et al.~\cite{acharyya12},
Ravikumar et al.~\cite{ravikumar2011ndcg} analysed the property of surrogate loss functions of normalised discounted cumulative gain (nDCG)~\cite{jarvelin2002cumulated} and proved that
any surrogate loss function that is strongly statistically consistent
to nDCG has the form of the Bregman divergence. 
In contrast to these LTR methods based on the Bregman divergence,
we employ a semi-supervised DRE approach for handling the one-class problem of implicit feedback.
Ren et al.~\cite{ren2011importance} proposed a weighted variant of AdaRank~\cite{xu2007adarank} in which
the importance weights (i.e. density ratio) are estimated through the
Kullback-Leibler Importance Estimation Procedure (KLIEP)~\citep{sugiyama2008direct}.
Li et al. also utilised KLIEP for query-level weighting in unsupervised transfer ranking~\cite{li2016effectiveness}.
These methods aim to estimate the importance weights for samples through DRE and do not utilise the DRE approach for estimating ranking.
In this paper, we formulate personalised top-$K$ ranking as a minimisation of the Bregman divergence.
We leverage estimated density ratios for weighing samples in the same spirit as the conventional methods such as Ren et al.~\cite{ren2011importance} and Li et al.~\cite{li2016effectiveness}.

\section{General Framework Based on Density Ratio Estimation}
In this section, we provide the generalised form of the risk function in our framework.
We first derive the risk of the WMF for implicit feedback from the viewpoint of semi-supervised learning.
Then, we demonstrate that personalised ranking from implicit feedback can be feasible by estimating the ratio between two probability densities.
Finally, we develop a weighted variant of the conventional risk of DRE, which is our generalised formulation for learning personalised ranking.

\subsection{Problem Formulation}
In a scenario of personalised recommendation,
for a set of users $\mathcal{U}$ and a set of items $\mathcal{I}$,
we observe $N$ i.i.d. user feedback logs $(u_1, i_1, y_1),\dots,(u_N, i_N, y_N)$ where $u \in \mathcal{U}$, $i \in \mathcal{I}$ and $y \in \{-1, +1\}$ denote user, item and class label, respectively.
However, owing to implicit feedback, we can observe only positive samples; that is, a data labelled as $y=+1$.
The goal of this study is, when given a user, to sort items in the order in which items are likely to interact with the user.
Throughout this study, we assume the sparsity of interactions; the number of logs $N$ is much smaller than the number of all possible user-item pairs, $|\mathcal{U}||\mathcal{I}|$.

\subsection{Pointwise Loss Functions\label{pu-pointwise}}
In this section, we first introduce classification-based pointwise loss functions.
For $(u, i, y)$, we define the traditional pointwise loss function of WMF~\cite{hu2008collaborative} as follows:
\begin{eqnarray}
    \mathcal{L}_{wmf}(g; u,i, y, w_{u,i}) = w_{u,i}\left(\mathds{1}\{y = +1\} - g(u,i)\right)^2 ,
\end{eqnarray}
where $g$ is a model to be trained, and $w_{u,i}$ is the sample-dependent weight for $u$ and $i$.
Then, we can derive the risk which WMF aims to minimise as follows.
\begin{align}
\notag
 &R_{wmf}\\ \notag
 &= \mathbb{E}\left[\mathcal{L}_{wmf}(g; u,i, y, w_{u,i})\right] \\\notag
 &= \pi^{+}\mathbb{E}_{+}\left[w_{u,i}^{+}(1-g(u,i))^2\right] + \pi^{-}\mathbb{E}_{-}\left[w_{u,i}^{-}(0-g(u,i))^2\right]\\ \notag
  &= \pi^{+}\mathbb{E}_{+}\left[w_{u,i}^{+}(1-g(u,i))^2\right] + \\ \notag
   &+ \mathbb{E}\left[w_{u,i}^{-}(0-g(u,i))^2\right] - \pi^{+}\mathbb{E}_{+}\left[w_{u,i}^{-}(0-g(u,i))^2\right] \\\label{no-sampling-risk}
    &= \pi^{+}\underbrace{\mathbb{E}_{+}\left[(w_{u,i}^{+}-w_{u,i}^{-})g(u,i)^2-2w_{u,i}^{+}g(u,i)\right]}_{\substack{\text{estimated with} \\ \text{positive samples}}}
    + \underbrace{\mathbb{E}\left[w_{u,i}^{-}g(u,i)^2\right]}_{\substack{\text{estimated with} \\ \text{all samples}}} + C ,
\end{align}
where $C$ is a constant value with respect to the training of $g$.
Here, $\pi^{+}=p(y=+1)$ and $\pi^{-}=p(y=-1)$ denote class priors, and let $\mathbb{E}_{+}$ and $\mathbb{E}_{-}$ be the expectations over $p(u,i|,y=+1)$ and $p(u,i|y=-1)$, respectively.
We also let $w_{u,i}^{+}$ and $w_{u,i}^{-}$ be the class-dependent weights.
A similar derivation can be found in the recent non-sampling approaches~\cite{chen2019efficient,chen2020efficient,chen2020jointly};
on the other hand, the theoretical background has been established as positive-unlabelled (PU) learning before~\cite{elkan2008learning,du2015convex}.
The key idea of PU learning is that the expected risk over the negative samples can be expressed by the difference between those over the unlabelled and positive ones (the second equation in Eq.~(\ref{no-sampling-risk})).

While we demonstrate the above risk based on the WMF by following the derivation in PU learning,
the empirical risk in the literature of the method based on the non-sampling approach~\cite{chen2019efficient,chen2019social,chen2020efficient,chen2020jointly} does not consider $\pi^{+}$ as the approach aim to reduce sampling procedures.
Therefore, the empirical risk of the non-sampling approach cannot be safely computed on the data wherein both users and items are sub-sampled; thus, they require infeasible sampling strategies for large-scale data that leverage all the users or items in a mini-batch.

\subsection{Learning to Rank through Density Ratio Estimation\label{ltr-dre}}
To create a top-$K$ recommendation result for user $u$ with the above classification-based approach,
we sort all items $i \in \mathcal{I}$ according to the order of estimated class probabilities $p(y=+1|u,i)$. 
Here, based on the Bayes rule $p(y=+1|u,i)=p(u,i|y=+1)p(y=+1)/p(u,i)$,
the constant value $p(y=+1)$ does not affect to the order of classification probabilities,
and therefore, estimating $r(u,i)=p(u,i|y=+1)/p(u,i)$ is sufficient.
Hence, the problem of estimating $p(y=+1|u,i)$ is overly general to achieve top-$K$ recommendation;
this is one reason why the method based on the pointwise approach can impair the ranking performance.
The task of estimating the ratio between two probability densities such as $r(u,i)=p(u,i|y=+1)/p(u,i)$ are known as \emph{DRE}.

Considering this, we first formulate a DRE task through the minimisation of the gap between the true and estimated density ratios.
To directly estimate the ratio of probability densities $r(u,i)$ from positive-unlabelled data,
we utilise the Bregman divergence~\cite{sugiyama2012density}, which is a wide class of divergences, to quantify the gap between two density ratios.
The Bregman divergence is defined as follows:
\begin{eqnarray}
  BR_f(t||\hat{t}) = f(t) - f(\hat{t}) - \partial f(\hat{t}) (t - \hat{t}) ,
\end{eqnarray}
where $f$ is a strictly convex function, and $t \in \mathbb{R}$ is a scalar.

For providing a scalable learning framework for large-scale data and complex models,
we formulate a DRE task as an empirical risk minimisation (ERM), in which we approximately minimise a risk through minimisation of the sample average of loss functions (i.e. empirical risk) on sub-sampled examples;
hence, it allows us to utilise mini-batch SGD seamlessly.
The Bregman divergence enables the empirical approximation of the expected divergence between true density ratio $r(\cdot)$ and estimated density ratio $\hat{r}(\cdot)$ with unlabelled and positive user-item pairs in the historical feedback logs.
The risk based on the Bregman divergence can be expressed as follows:
\begin{align}
\notag
  & R(\hat{r}) \\\notag
  & = \mathbb{E}_{p(u,i)}\left[f(r) - f(\hat{r}) - \partial f(\hat{r}) (r - \hat{r}) \right]\\ \notag
  & = \mathbb{E}_{p(u,i)}\left[- f(\hat{r}) + \partial f(\hat{r}) \hat{r}\right] - \mathbb{E}_{p(u,i)}\left[r \partial f(\hat{r}) \right] + \mathbb{E}_{p(u,i)}\left[ f(r) \right]\\ \label{bregman_risk}
  & = \underbrace{\mathbb{E}_{p(u,i)} \left[- f(\hat{r}) + \partial f(\hat{r}) \hat{r}\right]}_{\substack{\text{estimated with} \\ \text{all samples}}} - \underbrace{\mathbb{E}_{p(u,i|y=+1)} \left[ \partial f(\hat{r}) \right]}_{\substack{\text{estimated with} \\ \text{positive samples}}} + \tilde{C} ,
\end{align}
where $\tilde{C}=\mathbb{E}_{p(u,i)}\left[f(r)\right]$ is a constant term irrelevant to the model, and $\partial f(\hat{r})$ denotes the derivative of $f$.
In the second equation, we utilise $p(u,i)r(u,i)=p(u,i|y=+1)$ to derive the second term.
Based on Eq.~(\ref{bregman_risk}), through the minimisation of the Bregman divergence between true density ratio $r=p(u,i|y=+1)/p(u,i)$ and the model of density ratio $\hat{r}$,
we can achieve LTR based on implicit feedback.
However, this formulation lacks an explicit \emph{top-weighted} property which is an essential factor in top-$K$ recommendation tasks,
whereas DRE-based risks originally weigh samples with large values of density ratio~\cite{menon2016linking}.

\subsection{Weighted DRE Risk for Implicit Feedback\label{weighted-risk}}
In this section, we first define the generalised form of our proposed risk, namely, weighted DRE risk, which considers weights for samples, thereby enabling installation of a top-weighted property in the risk.
By following the risk of the WMF,
we consider the minimisation of a weighted risk based on the Bregman divergence that can be expressed as follows:
\begin{align*}
  R_w(\hat{r}) = \mathbb{E}_{p(u,i)}\left[w_{u,i}(f(r)-f(\hat{r})-\partial f(\hat{r})(r-\hat{r}))\right] ,
\end{align*}
From Eq.~(\ref{bregman_risk}), we can derive the risk that can be computed on positive-unlabelled data as follows:
\begin{align*}
  & R_w(\hat{r}) \\
  & = \mathbb{E}\left[w_{u,i}(f(r)-f(\hat{r})-\partial f(\hat{r})(r-\hat{r}))\right]\\
  & = \mathbb{E}\left[w_{u,i}(f(\hat{r})-\partial f(\hat{r})\hat{r})\right] - \mathbb{E}_{+}[w_{u,i}^{+}\partial f(\hat{r})] + \tilde{C} ,
\end{align*}
Then, we let $\ell_{+}(\hat{r})=-\partial f(\hat{r})$ and $\ell_{\pm}(\hat{r})=f(\hat{r})-\partial f(\hat{r})\hat{r}$ and obtain the risk for implicit feedback with class-dependent weights.
\begin{align}
  \notag
  & R_w(\hat{r}) \\\notag
  & = \pi^{+}\mathbb{E}_{+}\left[w_{u,i}^{-}\ell_{\pm}(\hat{r})\right] + \pi^{-}\mathbb{E}_{-}\left[w_{u,i}^{-}\ell_{\pm}(\hat{r})\right] + \mathbb{E}_{+}[w_{u,i}^{+}\ell_{+}(\hat{r})] + \tilde{C} \\\notag
  & = \pi^{+}\mathbb{E}_{+}\left[w_{u,i}^{-}\ell_{\pm}(\hat{r})\right] + \mathbb{E}\left[w_{u,i}^{-}\ell_{\pm}(\hat{r})\right] - \pi^{+}\mathbb{E}_{+}\left[w_{u,i}^{-}\ell_{\pm}(\hat{r})\right] \\ \notag
  & + \mathbb{E}_{+}[w_{u,i}^{+}\ell_{+}(\hat{r})] + \tilde{C} \\ \label{wbr-risk}
  & = \mathbb{E}_{+}\left[\pi^{+}(w_{u,i}^{+}-w_{u,i}^{-})\ell_{\pm}(\hat{r}) + w_{u,i}^{+}\ell_{+}(\hat{r}) \right] + \mathbb{E}\left[w_{u,i}^{-}\ell_{\pm}(\hat{r})\right] + \tilde{C} ,
\end{align}
To remove class-independent weights $w_{u,i}$, we utilised $\mathbb{E}\left[w_{u,i}\ell_{\pm}(\hat{r})\right]=\pi_{+}\mathbb{E}_{+}\left[w_{u,i}^{+}\ell_{\pm}(\hat{r})\right]+\pi_{-}\mathbb{E}_{-}\left[w_{u,i}^{-}\ell_{\pm}(\hat{r})\right]$ in the first equation.
In the second equation, we utilised the same derivation in PU learning to remove the expectation over $p(u,i|y=-1)$ that is infeasible in implicit settings.

\section{Specialisation for Top-K Personalised Recommendation}
\subsection{User-wise DRE for Personalised Recommendation\label{user-wise-optimisation}}
In personalised recommendation tasks,
we are concerned with the order of items for each user and do not require the comparable preference quantities of items across users; therefore, estimating $r(i|u)=p(i|u,y=+1)/p(i|u)$ is sufficient.
Furthermore, evaluation measures in personalised recommendation are typically computed as the average of a ranking measure for each user,
such as R@$K$ and nDCG@$K$; these measures assume that the distribution of users $p(u)$ is a uniform distribution in the evaluation.
Thus, when na\"{i}vely minimising the empirical risk based on the observed user feedback $\mathbb{E}_{p(u,i,y)}[\ell(\hat{r}(x),y)]$,
the model experiences \emph{covariate shift}~\cite{shimodaira2000improving}, as the empirical distribution of $p(u)$ is generally not a uniform distribution.
To overcome this problem, we propose a user-wise DRE approach that minimises the following risk:
\begin{eqnarray}
  R_{wu}(\hat{r}) &=& \mathbb{E}_{p(u)}\mathbb{E}_{p(i|u)}\left[w_{u,i}BR_f(r(i|u)||\hat{r}(i|u))\right] ,
\end{eqnarray}
In the training phase, for ensuring $p(u)$ as a uniform distribution, we can simply take the average of the loss for each user, when the sampling distribution of users in a mini-batch is uniform.
To this end, we consider another form of the risk as follows:
\begin{align}
  \notag
  & R_{wu}(\hat{r}) \\
  & = \mathbb{E}_{p(u)}\left[\mathbb{E}_{+}\left[\pi_u^{+}(w_{u,i}^{+}-w_{u,i}^{-})\ell_{\pm}(\hat{r}) + w_{u,i}^{+}\ell_{+}(\hat{r})\right] + \mathbb{E}_{p(i|u)}\left[w_{u,i}^{-}\ell_{\pm}(\hat{r})\right]\right],
\end{align}
where $\mathbb{E}_{+}$ denotes the expectation over $p(i|u,y=+1)$, and we omitted constant values.
Positive constant value $\pi_u^{+}=p(y=+1|u)$ is a user-dependent class prior and can be regarded as the activity of user $u$.
In practice, it can be estimated by computing the ratio of positive items for a user in all the items, $|\mathcal{I}_{u}^{+}|/|\mathcal{I}|$.

\subsection{Hard-Sample Weighting for Top-K Ranking\label{hard-sample-weighting}}
In this section, we describe the implementation of our weighting strategy for top-$K$ ranking problems 
that explicitly induces a top-weighted property in our weighted DRE risk.

The widely adopted weighting strategy for the pointiwise approach is based on frequency/popularity of items~\cite{he2016fast,liang2016modeling}:
\begin{eqnarray}
  \label{popularity_weight}
  \tilde{w}_{u,i}^{+} = 1,\,\,\, \tilde{w}_{u,i}^{-} = c_0 \frac{|\mathcal{U}_i^{+}|^{\alpha}}{\sum_{i \in  \mathcal{I}}|\mathcal{U}_i^{+}|^{\alpha}} ,
\end{eqnarray}
where $c_0$ is a positive constant that controls the overall importance of unobserved samples, and $\alpha$ is a parameter that controls the skewness of the weights.
This weighting strategy assumes that frequent items have been exposed and examined by users sufficiently, and, therefore, unobserved samples of those items are likely to be true negatives and can be penalised safely.
However, such a strategy is indirect for approximating the risk around important samples in top-$K$ ranking evaluation wherein only the items in the top of ranked lists are critical.

We propose a weighting strategy which is aware of the top-weighted property of measures. To this end, we first define class-dependent weights by leveraging the estimated density ratio as follows:
\begin{eqnarray}
  \label{rank_pos_weight}
  w_{u,i}^{+} = c_u^{+}\frac{1}{\hat{r}(i|u)},\,\,\, w_{u,i}^{-} = c_u^{-}\hat{r}(i|u),
\end{eqnarray}
where $c_u^{+}$ and $c_u^{-}$ are positive constants for each user.
Since ranked lists are generated according to the predictions of model $\hat{r}$,
positive items ranked lower and negative ones ranked higher by $\hat{r}(i|u)$ are likely to be false negatives/positives in top-$K$ ranking evaluation.
Therefore, we penalise these items distributed nearby the decision boundary of $\hat{r}$.
This concept is reminiscent of the conventional negative sampling strategies~\cite{zhang2013optimizing,rendle2014improving,IRGAN,AdvIR} in which the sampler tries to select difficult samples for the ranker.
As a remarkable distinction between our weighting strategy and conventional ones,
we consider weights also for positive samples to pull up those in a lower position of a ranked list.
Such a weighting strategy for positive samples may hurt the ranking performance when we need to deal with multiple relevance grades; it may prioritise to pull up marginally relevant items rather than highly relevant ones.
However, in implicit settings, we can assume the binary relevance for items.

Our weighting strategy is scalable as it does not require retention of two large matrices with the size of $|\mathcal{U}||\mathcal{I}|$ for non-uniform weights, $w_{u,i}^{+}$ and $w_{u,i}^{-}$,
when the estimated score $\hat{r}(i|u)$ can be obtained such by the inner product of the embeddings of $u$ and $i$.
In this case, the space complexity for maintaining the embeddings is $O((|\mathcal{U}|+|\mathcal{I}|)d)$ where $d$ is the dimensionality of embeddings.
Moreover, to obtain further training efficiency,
we determine weights for each sample on the fly in each training step
and thus can save extra inference and storage costs for weighting.
The detailed procedure is shown in Algorithm~\ref{alg:training}.

It is worth mentioning that such adaptive weighting may cause a bias in ERM through mini-batch SGD.
In addition, due to the class-dependent weights, the estimated density ratio $\hat{r}$ is generally inconsistent to true density ratio $r$ even with static weighting.
However, we empirically demonstrate its efficiency and effectiveness in Section~\ref{ablation-weighting} and Section~\ref{ablation-adaptive}.

\subsection{Ranking uLSIF Risk}
In this section, we describe an implementation of our proposed risk.
We adopt $f(t)=(t-1)^2/2$, which corresponds to the unconstrained least squared importance fitting (uLSIF)~\cite{kanamori2009least} in the context of DRE, and obtain a risk similar to that of the WMF.
\begin{align}
\notag
  & R_{wu} \\
  & = \mathbb{E}_{p(u)}\left[\mathbb{E}_{+}\left[\frac{\pi_u^{+}}{2}(w_{u,i}^{+}-w_{u,i}^{-})\hat{r}^2 - w_{u,i}^{+}\hat{r} \right] + \frac{1}{2}\mathbb{E}_{p(i|u)}\left[w_{u,i}^{-}\hat{r}^2\right]\right],
\end{align}
Although we can obtain various risks based on our generalised risk in Eq.~(\ref{wbr-risk}) by choosing $f$, we leave this extension as a future work.

As discussed in Section~\ref{ltr-dre}, our DRE-based approach solves a more appropriate task (i.e. DRE) than that of the conventional pointwise approach (i.e. classification).
The same argument is valid for weighting strategies;
it is reasonable to weigh samples in a ranked list for a user according to \emph{relative} magnitude of our ranking-aware weights defined in Eq.~(\ref{rank_pos_weight}).
To realise such relative weighting, we introduce a self-normalisation technique for estimating the weighted risk for each user.
As a result, we obtain the empirical approximation of $R_{wu}(\hat{r})$ as follows:
\begin{eqnarray}
  \label{rank-dre-loss}
  \hat{R}_{wu}(\hat{r}) &=& \frac{1}{|\mathcal{U}_{B}|}\sum_{u \in U_{B}}\left(\hat{R}_u^{+,1}-\hat{R}_u^{+,2}-\hat{R}_u^{+,3}+\hat{R}_u^{\pm}\right),
\end{eqnarray}
where
\begin{align}
  &\hat{R}_{u}^{+,1} = \frac{\pi_u^{+}}{2\sum_{i \in \mathcal{I}_{u}^{+}}w_{u,i}^{+}}\sum_{i \in \mathcal{I}_{u}^{+}}w_{u,i}^{+}\hat{r}(i|u)^{2} ,\\
  &\hat{R}_{u}^{+,2} = \frac{\pi_u^{+}}{2\sum_{i \in \mathcal{I}_{u}^{+}}w_{ui}^{-}}\sum_{i \in \mathcal{I}_{u}^{+}}w_{ui}^{-}\hat{r}(i|u)^{2} ,\\
  &\hat{R}_{u}^{+,3} = \frac{1}{\sum_{i \in \mathcal{I}_{u}^{+}}w_{u,i}^{+}}\sum_{i \in \mathcal{I}_{u}^{+}}w_{u,i}^{+}\hat{r}(i|u) ,\\
  &\hat{R}_u^{\pm} = \frac{1}{2\sum_{i \in \mathcal{I}_{B}}w_{u,i}^{-}}\sum_{i \in \mathcal{I}_{B}}w_{u,i}^{-}\hat{r}(i|u)^2 ,
\end{align}
where $I_{u}^{+}$ denotes the positive items for a user.
In $\hat{R}_{u}^{+,1}$, $\hat{R}_{u}^{+,2}$, $\hat{R}_{u}^{+,3}$, and $\hat{R}_{u}^{\pm}$, we utilised the self-normalisation technique; we divide each weighted average by the sum of weights.
It allows us to ignore the scale of weights in a ranked list and thus to remove the positive constants $c_u^{+}$ and $c_u^{-}$ introduced in Eq.~(\ref{rank_pos_weight}).
By combining this self-normalisation technique and user-wise DRE approach  described in Section~\ref{user-wise-optimisation},
we can weigh samples relatively in a ranked sub-list for a user and thereby enabling listwise weighting without excessive computational costs.

In practice, we minimise the empirically approximated risk with $L_2$ regularisation as follows:
\begin{eqnarray}
    \hat{r} = \argmin_{\hat{r}} \left( \hat{R}_{wu}(\hat{r}) + \lambda \mathcal{R}(\hat{r}) \right) ,
\end{eqnarray}
where $\mathcal{R}(\hat{r})$ is a $L_2$ regularisation term, and $\lambda$ is a hyper-parameter to control intensity of the regularisation.

\begin{algorithm}[t]
  \caption{Training Algorithm}
  \label{alg:training}
  \begin{algorithmic}[1]
    \REQUIRE{Users $\mathcal{U}$,\\A set of sets of positive items for users $\mathcal{P}=\{\mathcal{I}_u^{+} | u\in \mathcal{U}\}$}
    \ENSURE{Model $\hat{r}$}
    \STATE Initialise model $\hat{r}$
    \FOR{number of training steps}
      \STATE Sample $|\mathcal{U}_B|$ users from $\mathcal{U}$ uniformly without replacement and obtain $\mathcal{U}_B$
      \STATE Obtain $\mathcal{I}_B$ by $\mathcal{I}_B = \bigcup_{u \in \mathcal{U}_B}\mathcal{I}_u^{+}$
      \STATE Compute ${\bf U}_B=\{{\bf e}_u|u \in \mathcal{U}_B\}$ and ${\bf I}_B=\{{\bf e}_i|i \in \mathcal{I}_B\}$
      \STATE Compute $\hat{r}(u,i)$ by $\mathrm{softplus}({\bf U}_B{\bf I}_B^{\top})$
      \STATE Determine weights $w_{u,i}^{+}$ and $w_{u,i}^{-}$
      \STATE Compute gradient of $\hat{R}(\hat{r})$ on $\mathcal{U}_B \times \mathcal{I}_B$ and update model $\hat{r}$
      \ENDFOR
    \RETURN $\hat{r}$
  \end{algorithmic}
\end{algorithm}

\subsection{Efficient Sampling Strategy and Risk Correction\label{sampling-method}}
As our DRE-based framework relies on empirical approximation,
the sampling strategy for mini-batches is an important factor for the stability of the risk estimation in each training step.
In typical recommendation scenarios, the observed interactions for each user are limited or even scarce, particularly for light users.
Therefore, the estimation of the expectation over $p(i|u,y=+1)$ is challenging and unstable.
To alleviate this problem, we adopt a user-based sampling strategy.
We first uniformly sample $|\mathcal{U}_B|$ users and then sample all positive items for each user.
Consequently, the set of items in a mini-batch, $|\mathcal{I}_B|$, is obtained as the union of the sets of users' positive items.
Hence, our approach is \emph{not} one of the non-sampling approaches as it leverages sub-sampled users/items;
however, it still leverages the whole data by employing all possible user-item pairs in each mini-batch, $\mathcal{U}_B \times \mathcal{I}_B$.
The detailed procedure is described in Algorithm~\ref{alg:training}.

We are concerned with the space complexity of this sampling strategy because recent models based on neural networks are trained with graphics processing units (GPUs) that have a severe constraint on their memory space.
As the space complexity generally results from the number of user-item pairs in a mini-batch, 
the complexity of our strategy in a single training step is $O(|\mathcal{U}_B||\mathcal{I}_B|)$.
Since $|\mathcal{I}_B| < |\mathcal{I}|$ holds, 
our strategy allows us to utilise a more flexible mini-batch size $|\mathcal{U}_B|$ than that of the user-based non-sampling approach whose average-case complexity is $O(|\mathcal{U}_B||\mathcal{I}|)$.
In practice, we can expect that $|\mathcal{I}_B| \ll |\mathcal{I}|$ holds in many applications, because (1) the average number of positive items for a user is small due to the sparsity; and (2) the popularity distribution often follows a power-law~\cite{harald2011item}; therefore, there is much overlap among the sets of positive items for users.
Our strategy is advantageous particularly in applications wherein the number of items is much larger than that of users (e.g. SNS and E-commerce) when compared to the user-based non-sampling approach.

The above sampling strategy can introduce a bias to empirical approximation by non-uniformly selecting items;
items frequently associated with users as positive items are oversampled.
The probability that item $i$ is sampled in a mini-batch can be computed as follows:
\begin{eqnarray}
\label{sampling-probability}
  s(i) = \begin{cases}
    1-\frac{_{|\mathcal{U}_{i}^{\pm}|}C_{|\mathcal{U}_{B}|}}{_{|\mathcal{U}|}C_{|\mathcal{U}_{B}|}}, & \text{if } |\mathcal{U}_{i}^{\pm}| > |\mathcal{U}_{B}|, \\
    0, & \text{if } |\mathcal{U}_{i}^{\pm}| \leq |\mathcal{U}_{B}|,
    \end{cases}
\end{eqnarray}
where $|\mathcal{U}_{i}^{\pm}|$ is the set of users who have not yet interacted with item $i$.
By utilising this sampling probability, we can offset the bias introduced by our sampling strategy as follows:
\begin{align*}
  & R_{wus}(\hat{r}) \\
  & = \mathbb{E}_{+}\left[\pi^{+}(w_{u,i}^{+}-w_{u,i}^{-})\ell_{\pm}(\hat{r}) + w_{u,i}^{+}\ell_{+}(\hat{r}) \right] + \mathbb{E}_{p(i|u)}\left[w_{u,i}^{-}\ell_{\pm}(\hat{r})\right]\\ \label{is-risk}
  & = \mathbb{E}_{+}\left[\pi^{+}(w_{u,i}^{+}-w_{u,i}^{-})\ell_{\pm}(\hat{r}) + w_{u,i}^{+}\ell_{+}(\hat{r}) \right] + \mathbb{E}_{s(i)}\left[\frac{p(i|u)}{s(i)}w_{u,i}^{-}\ell_{\pm}(\hat{r})\right]
\end{align*}
In recommendation tasks, as all items are sorted for creating a ranked list for each user, $p(i|u)$ can be ignored safely.
Accordingly, we can obtain a corrected variant of $\mathcal{L}$ in Eq.~(\ref{rank-dre-loss}) by replacing the approximated risk for unlabelled samples, $\hat{R}_{u}^{\pm}(\cdot)$, with the following one:
\begin{eqnarray}
  \hat{R}_u^{\pm} &=& \frac{1}{2\sum_{i \in \mathcal{I}_{B}}\frac{w_{u,i}^{-}}{s(i)}}\sum_{i \in \mathcal{I}_{B}}\frac{w_{u,i}^{-}}{s(i)}\hat{r}(i|u)^2 .
\end{eqnarray}

\subsection{Non-Negative Risk Correction for Neural Collaborative Filtering\label{nn-risk-br}}
Several works have pointed out the unbounded problem of the risks for both PU learning and DRE~\cite{kiryo2017positive,kato2020non};
that is, the risk defined in Eq.~\ref{wbr-risk} can be minimised infinitely by maximising the score for positive samples (See the term $-w_{u,i}^{+}\hat{r}$).
Hence, na\"{i}ve minimisation of the risk with a complex model, such as neural networks,
can lead to severe overfitting.
Therefore, we borrow the risk correction technique proposed by Kato and Teshima~\cite{kato2020non} for utilising flexible and complex models.
We can obtain a variant of our uLSIF-based risk as follows:
\begin{align}
\notag
  & R_{wusnn}(\hat{r}) \\\notag
  &= \mathbb{E}_{p(u)}\Bigg[\mathbb{E}_{+}\left[\frac{\pi_u^{+}}{2}(w_{u,i}^{+}-w_{u,i}^{-})\hat{r}^2 - w_{u,i}^{+}\hat{r} \right]+\frac{1}{2\bar{D}}\mathbb{E}_{+}\left[w_{u,i}^{+}\hat{r}^2\right] \\
  &+ \left(\frac{1}{2}\mathbb{E}\left[w_{u,i}^{-}\hat{r}^2\right]-\frac{1}{2\bar{D}}\mathbb{E}_{+}\left[w_{u,i}^{+}\hat{r}^2\right]\right)_{+}\Bigg] ,
\end{align}
where $\bar{D}$ is a positive constant that indicates the upper bound of density ratio, and $(\cdot)_{+} \coloneqq max(\cdot, 0)$.
This correction is based on the fact that
a model may overfit when the inner term of $(\cdot)_{+}$ is negative.
The empirical approximation of $R_{wusnn}(\hat{r})$ can be expressed as follows:
\begin{align}
\notag
  &\hat{R}_{wusnn}(\hat{r})\\
  &= \frac{1}{|\mathcal{U}_{B}|}\sum_{u \in U_{B}}\left(\hat{R}_u^{+,1}-\hat{R}_u^{+,2}-\hat{R}_u^{+,3}+ \hat{R}_u^{cor}+\left(\hat{R}_u^{\pm}- \hat{R}_u^{cor}\right)_{+}\right),
\end{align}
where
\begin{align}
  &\hat{R}_u^{cor} = \frac{1}{2\bar{D}\sum_{i \in \mathcal{I}_{u}^{+}w_{u,i}^{+}}}\sum_{i \in \mathcal{I}_{u}^{+}}w_{u,i}^{+}\hat{r}(u,i)^2 ,
\end{align}

\subsection{DRE-based Graph Neural Collaborative Filtering (DREGN-CF)\label{dregn-cf}}
This section describes our neural network-based method based on DRE and GNNs, \emph{DREGN-CF}.
For efficiently retrieving top-$K$ items from an enormous amount of candidates in the database,
inner-product-based models, which predict a user's rating for an item through the inner product between the user and item embeddings, are beneficial~\cite{rendle2020neural}.
They are also suitable for our framework that utilises all possible user-item pairs in a mini-batch by exhaustively predicting the score of the resulting pairs.
Considering this, we adopt GNN-based models that are capable of learning powerful representations
while enabling efficient inference through inner product between user and item embeddings~\cite{NGCF,he2020lightgcn}.
Based on these models, we introduce an activation function to the prediction step:
\begin{eqnarray}
  \hat{r}(i|u) = \mathrm{softplus}(\bf{e}_u^{\top}\bf{e}_i)
\end{eqnarray}
where ${\bf e}_u$ and ${\bf e}_i$ are the output embeddings of $u$ and $i$ built by the model.
We adopt a softplus activation function as density ratio $r(i|u)=p(i|u,y=+1)/p(i|u)$, that is non-negative and not upper-bounded.
It should be noted that we can ignore the activation function and cut off extra inference/comparison costs in production systems
as a softplus function is monotonically increasing with respect to the value of inner  product ${\bf e}_u^{\top}{\bf e}_i$ and does not affect the predicted order of samples.

In this study,
we do not focus on the architecture of neural networks to fairly examine our DRE-based framework and adopt the architecture of LightGCN~\cite{he2020lightgcn}.

\section{Experiments\label{expr}}
\subsection{Experimental Settings}
\subsubsection{Datasets}
To conduct fair comparison,
we closely follow the settings of the experiments in NGCF~\cite{NGCF} and LightGCN~\cite{he2020lightgcn}.
We utilise the same datasets and train/test splits that are available in GitHub\footnote{\label{repo-lightgcn}\url{https://github.com/kuandeng/LightGCN}};
(a) Gowalla~\cite{liang2016modeling}; (b) Yelp2018~\footnote{\url{https://www.yelp.com/dataset/challenge}}; and (c) Amazon-Book~\cite{he2016ups}.
The statistics are shown in Table~\ref{table:statistics}.

\begin{table}[t]
\caption{Statistics of the four datasets: Gowalla, Yelp2018, Amazon-Book.}
  \centering
\begin{tabular}{lrrrr}
  \hline
  \multicolumn{1}{l}{Dataset} & \multicolumn{1}{r}{User \#} & \multicolumn{1}{r}{Item \#} & \multicolumn{1}{r}{Interaction \#} & \multicolumn{1}{r}{Density} \\ \hline
  Gowalla & $29,858$ & $40,981$ & $1,027,370$ & $0.00084$ \\ 
  Yelp2018 & $31,668$ & $38,048$ & $1,561,406$ & $0.00130$ \\ 
  Amazon-Book & $52,643$ & $91,599$ & $2,984,108$ & $0.00062$ \\ \hline
    \end{tabular}
  \label{table:statistics}
\end{table}

\subsubsection{Baselines and Model Settings}
We list state-of-the-art methods based on neural networks.
\begin{itemize}
\item\textbf{Neural Graph Collaborative Filtering (NGCF)}~\cite{NGCF} This is a GNN-based method that inherits most of the design of graph convolutional network~\cite{GCN}. 
\item\textbf{Multi-VAE}~\cite{liang2018variational} This is a VAE-based collaborative filtering method. This method can be considered as a listwise approach.
\item\textbf{Efficient Neural Matrix Factorisation (ENMF)}~\cite{chen2019efficient} This method adopts a model based on neural matrix factorisation. The model architecture is originally proposed in the work of Neural Collaborative Filtering (NCF)~\cite{NCF}. It adopts a user-based non-sampling approach.
\item\textbf{LightGCN}~\cite{he2020lightgcn} This is a state-of-the-art GNN-based method and has been shown to outperform NGCF and Multi-VAE~\cite{he2020lightgcn}. It can handle higher order relationships on the graph of users and items. It adopts the pairwise approach.

\end{itemize}
We utilise the implementation of each method released by the authors in GitHub for these four methods; NGCF~\footnote{\url{https://github.com/xiangwang1223/neural_graph_collaborative_filtering.git}}, Multi-VAE~\footnote{\url{https://github.com/dawenl/vae_cf}}, ENMF~\footnote{\url{https://github.com/chenchongthu/ENMF}}, and LightGCN~\footnotemark[\getrefnumber{repo-lightgcn}]\footnote{\url{https://github.com/gusye1234/LightGCN-PyTorch}}.
For NGCF, ENMF and LightGCN, we utilise the best parameters reported by authors for each dataset, when those are reported.
For Multi-VAE, we utilised the model architecture suggested one in the paper (600 $\rightarrow$ 200 $\rightarrow$ 600).
We set the dimensionality of the embeddings to 64 for ENMF.

To keep comparisons fair,
our DREGN-CF utilises the same network architecture as LightGCN; throughout this study, in LightGCN-based methods including our DREGN-CF, we set the number of the light-weight graph convolution layers and the dimensionality of embeddings to 3 and 64, respectively.  
For our DREGN-CF, we have two hyper-parameters, namely, the weight for $L_2$ regularisation $\lambda$ and the constant value for non-negative risk correction $\bar{D}$.
We search the values of $\lambda$ and $\bar{D}$ based on validation within the range of $\{0.01, 0.02, \cdots, 0.09\}$ and $\{10, 20,\cdots, 90\}$.
We set the mini-batch size to $9500$, $12000$, and $6000$ for Gowalla, Yelp2018, and Amazon-Book, respectively; these values are determined by the maximum size that can be stored in the 16GB VRAM of a NVIDIA Tesla P100.
The learning rate is tuned in $\{0.01, 0.02, \cdots, 0.1\}$ for all the dataset.
The class prior for each user $\pi_u^{+}$ is estimated on the training split by $\pi_u^{+}=|\mathcal{I}_{u}^{+}|/|\mathcal{I}|$.
We divided the original training split utilised in the works of NGCF and LightGCN into training and validation splits with the ratio of 9:1.
We determine all the hyper-parameters on the validation split.

\subsection{Comparison of Effectiveness of Methods\label{expr-effectiveness}}
We first evaluate our proposed method, DREGN-CF, and baselines in terms of overall ranking effectiveness.
By following the conventional works~\cite{NGCF,he2020lightgcn},
we utilise R@20 and nDCG@20 as the evaluation measures throughout this study.
Table~\ref{table:overall_effectiveness} demonstrates R@20 and nDCG@20 of each method in the three datasets.

In Yelp2018 and Amazon-Book datasets, our DREGN-CF outperforms all methods with a substantial margin in terms of both R@20 and nDCG@20.
In Gowalla dataset, it outperforms the other methods except LightGCN; however, the difference between DREGN-CF and LightGCN is not substantial.
The gain from LightGCN in Yelp2018 and Amazon-book is remarkable despite the model architecture is the same; the main difference of these methods is only in the optimisation approach.
These results suggest that our DRE-based approach can be an option for top-$K$ ranking problems in terms of ranking effectiveness
as it does not require any complicated sampling strategies and any modifications in the model architecture.

It is worth mentioning that the top-weighted property of each method can be observed more clearly in Amazon-Book dataset;
such a property takes on more importance when the pool depth of items is large (See also Table~\ref{table:statistics}).
Multi-VAE shows comparable performance to that of LightGCN in the dataset.
It suggests that its multinomial likelihood-based approach successfully induces a top-weighted property into the model.
By contrast, ENMF, which is based on the pointwise approach, significantly deteriorate in the dataset.
Our DREGN-CF shows a large gain particularly in Amazon-Book dataset; it indicates that our ranking uLSIF risk with hard-sample weighting is strongly top-weighted.
It should be noted that ENMF adopts a uniform weighting strategy for unobserved samples, and it can be one of the cause of the performance degradation in Amazon-Book dataset.
We discuss the effect of weighting strategies on the ranking effectiveness in Section~\ref{ablation-weighting}.

\begin{table}[t]
  \caption{Comparison of methods in terms of overall ranking effectiveness.}
  \label{table:overall_effectiveness}
  \resizebox{0.47\textwidth}{!}{
    \begin{tabular}{l|cc|cc|cc}
      \hline
      \multirow{2}{*}{Method} & \multicolumn{2}{c|}{Gowalla} & \multicolumn{2}{c|}{Yelp2018} & \multicolumn{2}{c}{Amazon-Book} \\ \cline{2-7}
                & R@20   & nDCG@20& R@20   & nDCG@20& R@20   & nDCG@20 \\ \hline
      NGCF      & 0.1567 & 0.1325 & 0.0575 & 0.0474 & 0.0342 & 0.0262 \\ 
      Mult-VAE  & 0.1644 & 0.1340 & 0.0589 & 0.0458 & 0.0413 & 0.0306 \\ 
      ENMF      & 0.1512 & 0.1306 & 0.0622 & 0.0513 & 0.0344 & 0.0272 \\ 
      LightGCN  & 0.1828 & 0.1551 & 0.0651 & 0.0532 & 0.0421 & 0.0324 \\ \hline
      DREGN-CF  & 0.1810 & 0.1541 & 0.0687 & 0.0564 & 0.0482 & 0.0378 \\ \hline
    \end{tabular}}
\end{table}

\subsection{Comparison of Training Efficiency of Methods\label{expr-efficiency}}
In this section, we evaluate our proposed method and baselines in terms of training efficiency, such as the number of training iterations needed to converge.
To fairly evaluate the efficiency of methods with different implementations,
we compare methods in terms of both run time and the number of training iterations.
We run the experiments in this section with one NVIDIA Tesla P100 GPU, 16 Intel(R) Xeon(R) CPU @ 2.20GHz, and 60GB of RAM.

\begin{figure}[t!]
    \centering
    \includegraphics[clip,width=\linewidth]{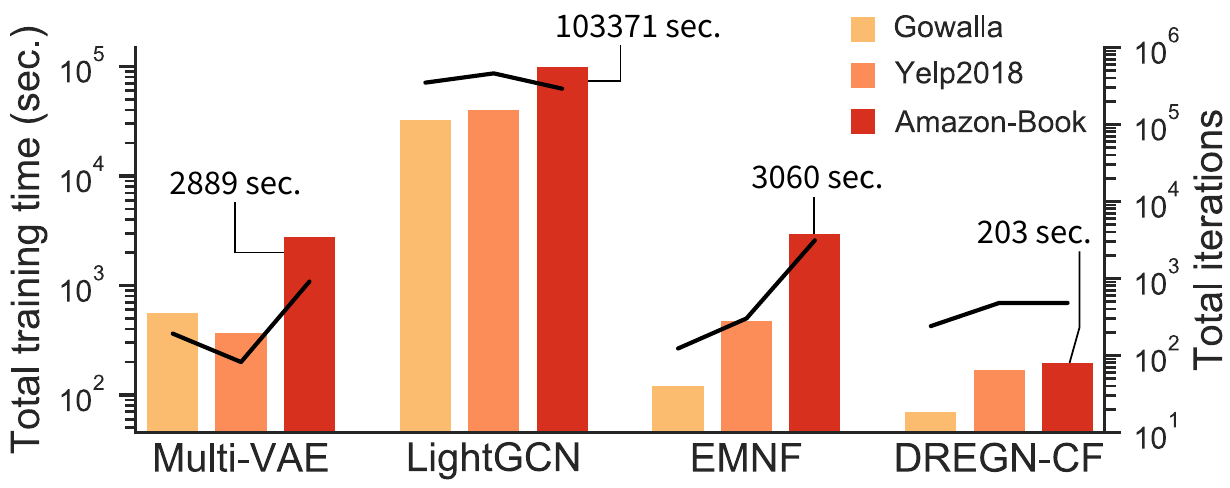}
    \caption{Growth of training time for different datasets. The bar plots indicate the total time shown in seconds. The line plots indicate the number of total training iterations.}
    \label{fig:training-efficiency}
\end{figure}

Figure~\ref{fig:training-efficiency} demonstrates the training efficiency of each method on the three datasets.
The bar and line plots shown for each method indicate the total run time and number of iterations to converge; the corresponding y-axes are shown in the left and right sides in the figure, respectively.
The y-axes are shown in logarithmic scales.

Our DREGN-CF demonstrates significant improvement from LightGCN in terms of the number of iterations and training run time to converge.
As these two methods have the same model architecture, this gain is the result of our DRE-based approach.
On the other hand, DREGN-CF requires more training iterations to converge than ENMF in Gowalla and Yelp2018 datasets (See the line plots in Figure~\ref{fig:training-efficiency});
ENMF only takes 123 and 302 training iterations for Gowalla and Yelp2018 datasets while DREGN-CF requires 240 and 480 iterations.
However, as ENMF utilises all the items in a training iteration, its consumption time for each iteration is much longer than that of DREGN-CF; ENMF requires 1.036 and 1.630 seconds for each iteration in Gowalla and Yelp2019 datasets while DREGN-CF only takes 0.301 and 0.360 seconds.
Thus, the total training time of ENMF is longer than DREGN-CF (the bar plots in the Figure~\ref{fig:training-efficiency}).
For the Amazon-Book dataset, which has the largest size of items among the three datasets,
our DREGN-CF significantly outperforms ENMF in terms of all the criteria, namely, the total training time, consumption time for a single training iteration, and, total number of iterations.
In particular, ENMF requires 3,120 training iterations in Amazon-Book dataset whereas our DREGN-CF takes only 480 iterations.
As discussed in Section~\ref{sampling-method}, the user-based non-sampling approach (e.g. ENMF) cannot utilise a large mini-batch size (i.e. the number of users) due to the space complexity that results from using all the items in a dataset for each training step. 
As a result, ENMF significantly deteriorates in terms of training efficiency in Amazon-Book dataset.
On the other hand, Multi-VAE shows a similar trend to that of ENMF; it takes 2,889 seconds of total training time and 910 iterations to converge for the Amazon-Book dataset. In particular, its average consumption time for a single iteration reaches 3.156 seconds in the dataset.
This is because Multi-VAE requires to compute the multinomial probability across all the items for the normalisation of probabilities;
however, this issue may be alleviated by approximating the normalisation factor~\cite{botev2017complementary} as suggested by Liang et al.~\cite{liang2018variational}.
By contrast, our DREGN-CF demonstrates its scalability against the number of items in the datasets.

\begin{table*}[t]
  \caption{Comparison of overall ranking effectiveness among ablated methods. The rows indicate (a) LightGCN with weighted PU regression, (b) LightGCN with weighted DRE, (c) DREGN-CF without importance sampling for removing sampling bias, (d) DREGN-CF without non-negative risk correction, (e) DREGN-CF with uniform sample weighting, (d) DREGN-CF with popularity-based weighting, (g) DREGN-CF with static hard-sample weighting, and (h) DREGN-CF.}
  \label{table:ablation}
  \resizebox{1.0\textwidth}{!}{
    \begin{tabular}{lcccccc|cccccc}
\hline
\multirow{2}{*}{Method} & \multirow{2}{*}{DRE} & \multirow{2}{*}{\begin{tabular}[c]{@{}l@{}}ranking-\\ DRE\end{tabular}} & \multirow{2}{*}{IS} & \multirow{2}{*}{NN} & \multirow{2}{*}{\begin{tabular}[c]{@{}l@{}}adaptive-\\ weighting\end{tabular}} & \multirow{2}{*}{weight} & \multicolumn{2}{c}{Gowalla} & \multicolumn{2}{c}{Yelp2018} & \multicolumn{2}{c}{Amazon-Book} \\ \cline{8-13} 
                        &                      &                                                                         &                     &                     &                                                                                &                         & R@20        & nDCG@20       & R@20        & nDCG@20        & R@20          & nDCG@20         \\ \hline
      (a) PU		&      &      &      &      &      & uniform     & 0.1021 & 0.1232 &  0.0451 &  0.0382 & 0.0292 & 0.0232	\\ 
      (b) DRE		& \checkmark & 	  &      & 	    &      & uniform	  & 0.1024 & 0.1327 & 0.0472 & 0.0407 & 0.0344 & 0.0263 \\ 
      (c) w/o IS	& \checkmark & \checkmark &      & \checkmark & \checkmark & hard-sample & 0.1808 & 0.1542 & 0.0689 & 0.0566 & 0.0496 & 0.0386 \\ 
      (d) w/o NN	& \checkmark & \checkmark & \checkmark & 	    & \checkmark & hard-sample & 0.1798 & 0.1537 & 0.0683 & 0.0563 & 0.0481 & 0.0375 \\ 
      (e) Uni-W		& \checkmark & \checkmark & \checkmark & \checkmark &      & uniform	  & 0.1549 & 0.1324 & 0.0635 & 0.0525 & 0.0333 & 0.0260 \\ 
      (f) Pop-W		& \checkmark & \checkmark & \checkmark & \checkmark &      & popularity  & 0.1575 & 0.1347 & 0.0637 & 0.0526 &  0.0452 &  0.0361 \\ 
      (g) Static	& \checkmark & \checkmark & \checkmark & \checkmark &      & hard-sample & 0.1176 & 0.0955 & 0.0273 & 0.0222 & 0.0160 & 0.0120 \\ \hline
      (h) Full		& \checkmark & \checkmark & \checkmark & \checkmark & \checkmark & hard-sample & 0.1810 & 0.1541 & 0.0687 & 0.0564 & 0.0482 & 0.0378 \\ \hline
    \end{tabular}
  }
\end{table*}

\subsection{Ablation Study and Discussion}
Our framework has four primary distinctions from the conventional methods;
(1) DRE-based risk;
(2) risk correction for the user-wise sampling strategy;
(3) ranking-aware sample weighting; and
(4) adaptive weighting strategy.
To provide insightful discussion about the property of our framework,
we extensively examine each module in the above list.
As in the previous experiments, we utilise LightGCN as the base model.
The results are shown in Table~\ref{table:ablation}.

\subsubsection{PU Regression vs. DRE}
To validate the effectiveness of our DRE-based approach,
we compare the simplest losses based on our weighted DRE risk and weighted PU regression risk, which can be considered as the expected loss in the conventional WMF ~\cite{hu2008collaborative} as was discussed in Section~\ref{ltr-dre}.
The empirical approximation of PU in Eq.~(\ref{no-sampling-risk}) and uLSIF risks in Eq.~(\ref{wbr-risk}) can be expressed as follows:
\begin{align}
    \notag
    & \hat{R}_{pu} \\
    &= \sum_{u \in U_{B}}\left(\frac{\pi_{u}^{+}}{|\mathcal{I}_u^{+}|}\sum_{i \in \mathcal{I}_u^{+}}\left(\left(w_{u,i}^{+}-w_{u,i}^{-}\right)\hat{r}^2-2w_{u,i}^{+}\hat{r}\right)+\sum_{i \in \mathcal{I}_B}\frac{w_{u,i}^{-}}{{|\mathcal{I}_B|}}\hat{r}^2 \right) \\ \notag
    & \hat{R}_{wu} \\
    &= \sum_{u \in U_{B}}\left(\frac{1}{|\mathcal{I}_u^{+}|}\sum_{i \in \mathcal{I}_u^{+}}\left(\pi_{u}^{+}\left(w_{u,i}^{+}-w_{u,i}^{-}\right)\hat{r}^2-2w_{u,i}^{+}\hat{r}\right)+\sum_{i \in \mathcal{I}_B}\frac{w_{u,i}^{-}}{{|\mathcal{I}_B|}}\hat{r}^2 \right) 
\end{align}
In this section, we utilise the same uniform weighting strategy for each method; moreover, for fair comparison, we omit the non-negative correction for the DRE risk.
We also utilise the same $L_2$ regularisation for both methods.
The results of methods based on PU regression and DRE risks are shown as (a) and (b) in Table~\ref{table:ablation}, respectively.
By comparing the results of (a) and (b), the DRE-based method achieves consistently higher performance than the PU regression counterpart for the three dataset; these results are consistent to those demonstrated in Menon et al.~\cite{menon2016linking}.
However, the method based on the na\"{i}ve DRE-based risk (i.e. uLSIF) without our ranking-aware weighting strategy still underperforms that based on the pairwise ranking loss~\cite{rendle2009bpr}; this is the original loss function of LightGCN (See Table~\ref{table:overall_effectiveness}).

\subsubsection{Risk Correction Techniques}
This section investigates the effects of our risk correction techniques in Section~\ref{sampling-method} and Section~\ref{nn-risk-br}.
In Table~\ref{table:ablation}, row (c) shows the performance of DREGN-CF without the risk correction for removing sampling bias for items defined in Eq.~(\ref{is-risk}).
Remarkably, the method without the correction consistently outperforms our full method (h).
As sampling probability $s(i)$ is monotonically increasing with respect to the empirical popularity of item $i$ (see also Eq.~(\ref{sampling-probability})),
the method without the correction penalises unlabelled samples with popular items more than those with unpopular ones;
this leads to a similar effect to that of the popularity-based sampling strategy~\cite{rendle2014improving} which oversamples negative items with high frequency.
Therefore, this result aligns with the prior results; the popularity-based sampling strategy can improve top-$K$ ranking performance~\cite{zhang2013optimizing,rendle2014improving}.
Therefore, it also suggests that our framework can be further improved by utilising advanced adaptive sampling strategies~\cite{zhang2013optimizing,rendle2014improving}.

The result of our method without non-negative correction described in Section~\ref{sampling-method} can be shown as (d) ``w/o NN'' in the Table~\ref{table:ablation}.
In all the dataset, the ablated variant slightly underperforms our full method (h);
however, as the differences are not substantial, we recommend to use it when the model is highly complex, or it can be omitted when the training efficiency is the main concern.

\subsubsection{Weighting Strategy\label{ablation-weighting}}
In this section, we examine the weighting strategies in our weighted DRE risk described in Section~\ref{hard-sample-weighting}.
We consider two weighting strategies as baselines;
(e) uniform weighting; and (f) frequency-based weighting.
In (f), we weigh the positive samples uniformly by 1 and unlabelled samples uniformly by the constant value $c_0$.
The method of (e) weighs positive samples uniformly by 1 and unlabelled samples non-uniformly based on the popularity of items (See also Eq.~(\ref{popularity_weight})).
By comparing (e) and (f), the method with popularity-based weighting (f) consistently outperforms that with uniform weighting (e).
Moreover, the performance of the method with popularity-based weighting is competitive with the conventional state-of-the-art methods (See also Table~\ref{table:overall_effectiveness}).
This result indicates the effectiveness of our loss function based on DRE.
The performance gain obtained by popularity-based weighting is relatively large particularly in Amazon-Book dataset.
This result suggests that non-uniform weighting is essential for training models on datasets with large sets of items.
The method with our hard-sample weighting strategy (h) outperforms the other methods with a large margin for all the datasets; this result is consistent with that obtained in the negative sampling strategies~\cite{zhang2013optimizing,rendle2014improving}.
Moreover, our weighting strategy shows substantial gain consistently in the three datasets, whereas
the performance of the popularity-based strategy is not stable; the popularity-based strategy can be inappropriate for some applications as it relies on the empirical belief that the missing samples of popular items are true negatives.
In addition, we observed that the method with popularity-based strategy is quite sensitive to hyper-parameter $\alpha$ in Eq.~(\ref{popularity_weight}).
By contrast, our hard-sample weighting strategy is hyper-parameter-free when utilised with the self-normalisation technique in Eq.~(\ref{rank-dre-loss}).

\subsubsection{Static Updating vs. Adaptive Updating of Weights\label{ablation-adaptive}}
Although our adaptive weighting strategy may impair the unbiasedness of our ERM-based stochastic optimisation, it is highly effective as we demonstrated in Section~\ref{ablation-weighting}.
To validate our adaptive weighting strategy, we consider a non-adaptive variant of our strategy as (g), in which we train an initial model based on uniform weighting, then train the final model with weighting samples based on the density ratio estimated by the initial model; the weights for samples are fixed in the training process of the final model.

\begin{figure}[t!]
    \centering
    \includegraphics[clip,width=\linewidth]{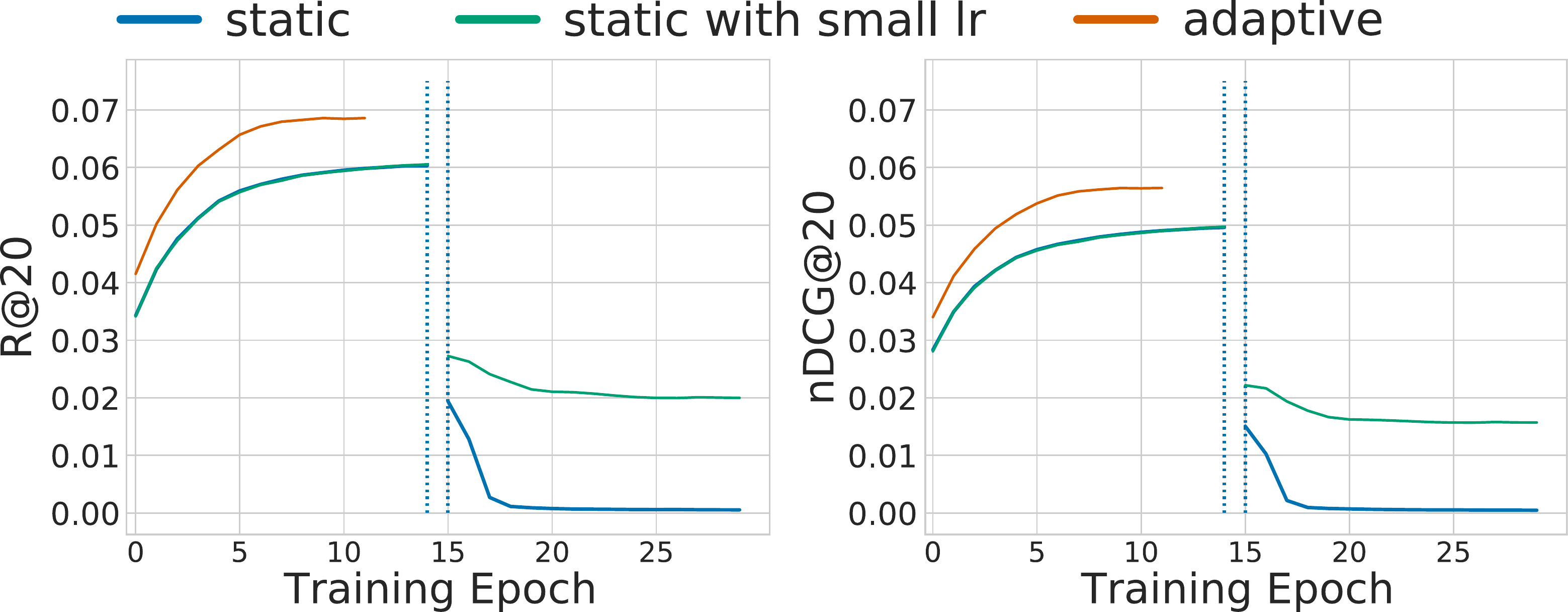}
    \caption{Training process of methods with static and adaptive weighting strategies. The left and right dotted lines in each figure indicate the last epoch of the model for weighting and the first epoch of the final model in the static strategies, respectively.}
    \label{fig:static-vs-adaptive}
\end{figure}

We observed that the static weighting strategy collapse training, as shown in the performance drop of (g) from the base model (e).
Figure~\ref{fig:static-vs-adaptive} shows the training curve of the methods with static and adaptive weighting strategies on Yelp2018 dataset.
The x-axis indicates the training epoch, and the y-axes are R@20 for the left and nDCG@20 for the right, respectively.
The orange line shows our adaptive strategy, the blue and green lines are the static strategies; the blue line is the strategy with the same learning rate for the initial and final models, and the method of the green line utilises a tenth part of the learning rate of the initial model for the final model training.
The initial model for static strategies is the model with the best hyper-parameter setting on the validation split.
The batch size is the same for all the methods.

The methods with static strategies (blue and green) show rapid deterioration in the early phase.
Particularly, the method with same learning rate fails to train the final model even in the later phase.    
Moreover, even by decreasing learning rate significantly, the static weighting strategy cannot achieve a higher performance in the final model than that in the first model.
By contrast, The method with our adaptive weighting strategy demonstrates rapid improvement and stable training process.

These results indicate, our adaptive weighting strategy is vital for our framework.
In addition, it does not require extra inference, storage costs, and multiple models as it generates weights by itself on the fly.

\subsection{Comparison of LightGCN with Dynamic Negative Sampling and DREGN-CF}
As discussed in Section~\ref{hard-sample-weighting},
our hard-sample weighting strategy is inspired by adaptive negative sampling strategies~\cite{zhang2013optimizing,rendle2014improving,IRGAN},
which prioritises informative samples by regarding the status of a model.
To validate the advantages of our weighting strategy, we consider three baseline methods for our DREGN-CF;
(1) LightGCN (pairwise ranking loss with uniform negative sampling);
(2) DREGN-CF-UNI-W (ranking uLSIF risk with uniform weighting); and
(3) LightGCN-DNS (pairwise ranking loss with adaptive negative sampling).
DREGN-CF-UNI-W is a variant of DREGN-CF with uniform weighting, which has been mentioned in Section~\ref{ablation-weighting} (See also row (e) in Table~\ref{table:ablation}).
For the method (3), we adopt dynamic negative sampling (DNS)~\cite{zhang2013optimizing}, which is known as one of the state-of-the-art adaptive samplers
and shows empirically strong performance in various datasets and tasks~\cite{rendle2014improving,ding2019rein,wang2020reinforced}.
DNS selects an unlabelled sample with a relatively high score among the candidates and rejects the other samples.
It should be noted that we do not utilise any adaptive sampling methods for our DREGN-CF to examine the effect of our hard-sample weighting strategy.

\begin{figure}[t]
    \centering
    \includegraphics[clip,width=\linewidth]{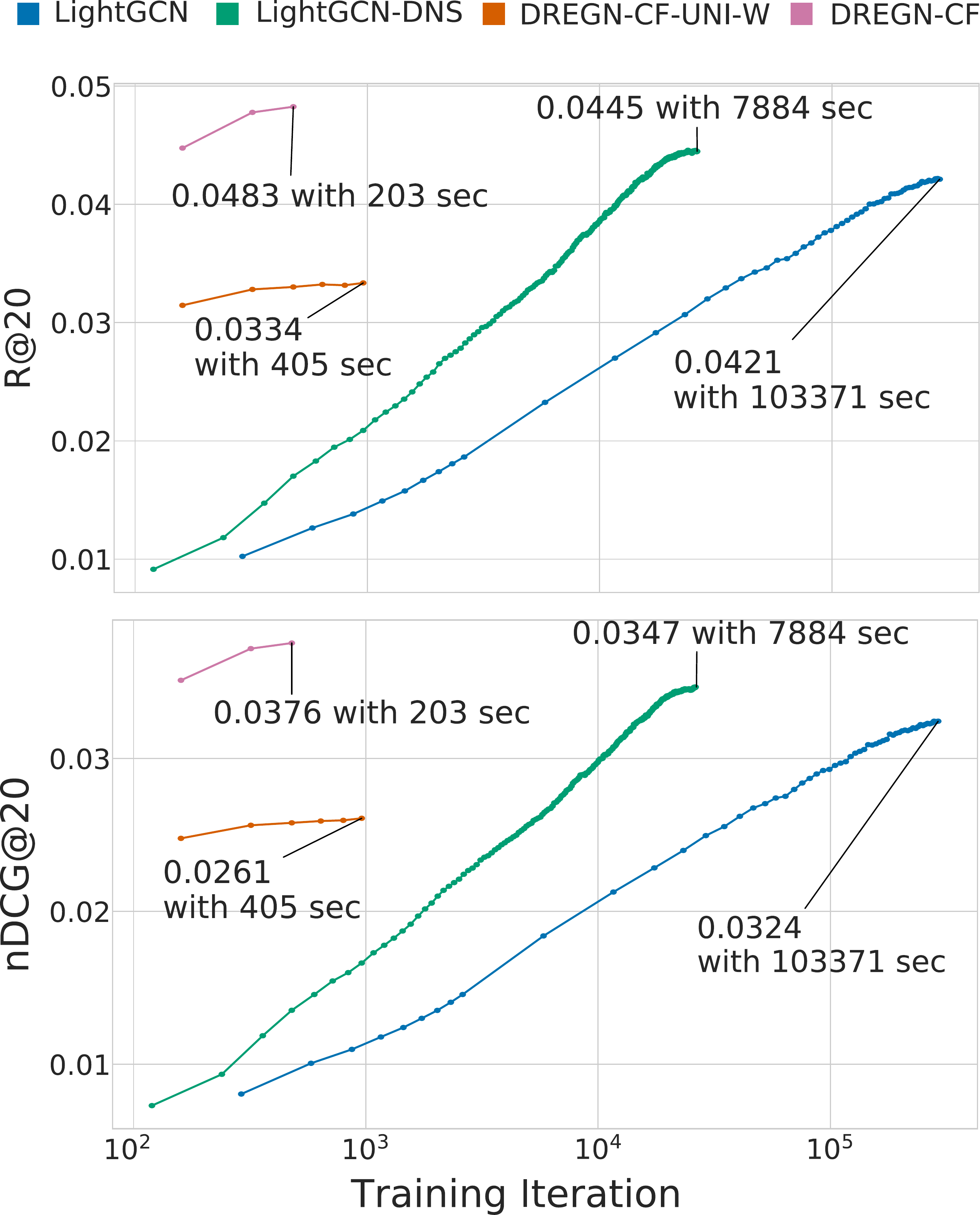}
    \caption{Training curves of the varints of LightGCN, LightGCN, LightGCN-DNS, DREGN-CF-UNI-W, and DREGN-CF, which are evaluated by testing R@20 and nDCG@20 on Amazon-Book dataset.}
    \label{fig:vs-dns-training-curve}
\end{figure}

Figure~\ref{fig:vs-dns-training-curve} demonstrates the training curves of LightGCN, LigtGCN-DNS, DREGN-CF-UNI-W, and DREGN-CF on Amazon-Book dataset. 
The x-axes indicate the number of training iterations, and the y-axes indicate the ranking performance in the test split.
The x-axis is shown in a logarithmic scale.
By comparing LightGCN (blue) and LightGCN-DNS (green), we can observe that DNS reduces the convergence time significantly while enhancing the final ranking performance.
DREGN-CF-UNI-W (orange) underperforms all the methods in terms of ranking performance whereas it shows rapid convergence.
On the other hand, our DREGN-CF (pink) achieves significant improvement even from LightGCN-DNS (green) with a large margin in terms of both ranking performance and convergence time.
In addition, DREGN-CF shows faster convergence than DREGN-CF-UNI-W.  

There are two observations from these results; (a) most of the gain in the training efficiency is obtained by our DRE-based pointwise approach
whereas our hard-sample weighting strategy enables further speeding up; and
(b) the weighting strategy mainly contributes to the final ranking performance.
The result of (a) indicates that 
our weighting strategy can efficiently leverage a large amount of samples including the unlabelled ones in a mini-batch by appropriately weighting them.
On the other hand, LightGCN-DNS is still inefficient due to the pairwise ranking optimisation and the rejection of negative samples in the sampling process.
The result of (b) indicates that,
the hard-sample weighting strategy successfully induces a top-weighted property and improves the top-$K$ ranking performance.

\subsection{Hyper-Parameter Sensitivity}
This section reports the sensitivity of DREGN-CF to the hyper-parameters;
(1) the coefficient for $L_2$ regularisation $\lambda$; and
(2) the constant value for non-negative correction, $\bar{D}$.
We vary $\lambda$ from 0.01 to 0.09 and $\bar{D}$ from $10$ to $90$, while keeping other parameters fixed.

\begin{figure}[ht]
    \centering
    \includegraphics[height=4.0cm]{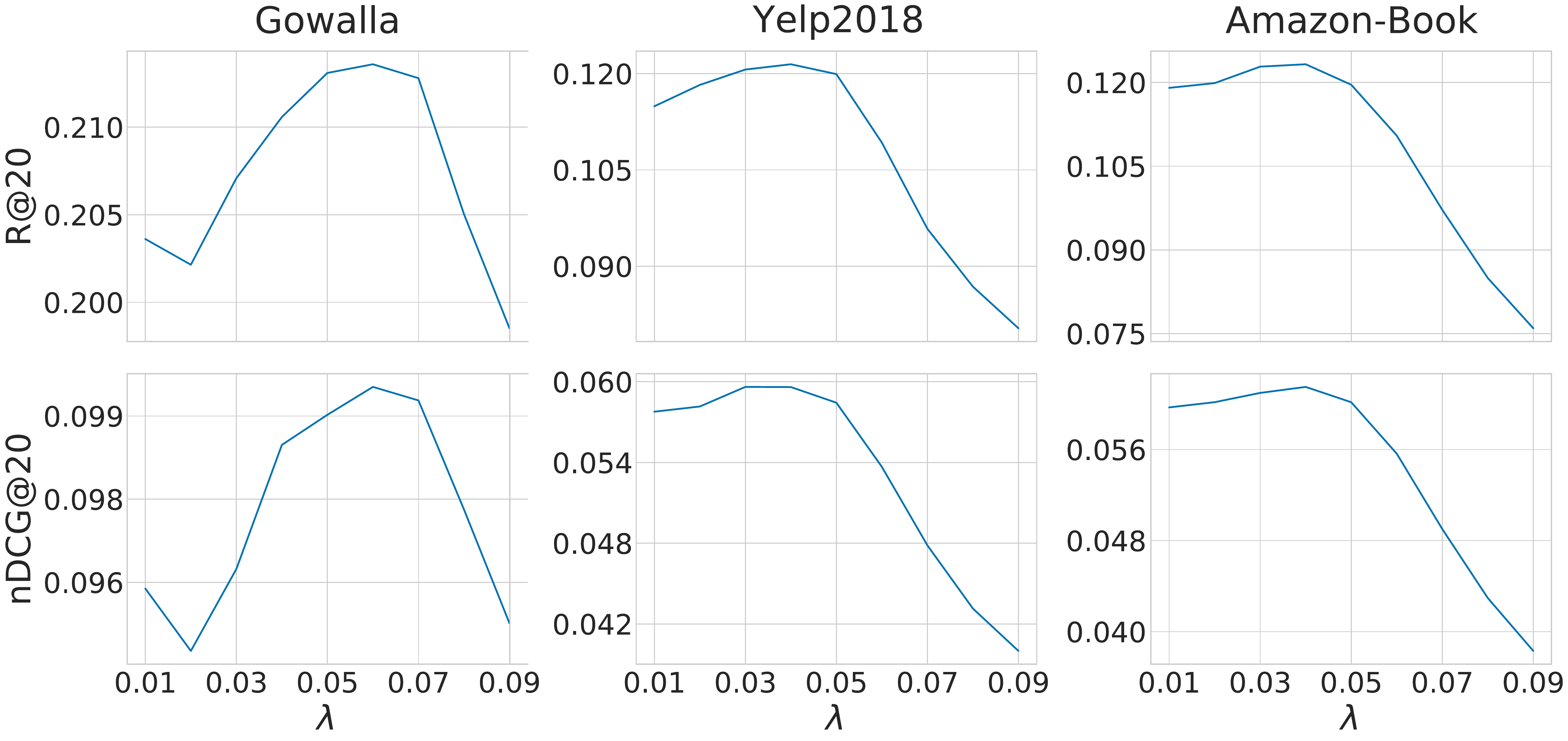}
    \caption{Effect of $\lambda$ on R@20 and nDCG@20 on the validation split.}
    \label{fig:sensitivity_lambda}
\end{figure}

\begin{figure}[ht]
    \centering
    \includegraphics[height=4.0cm]{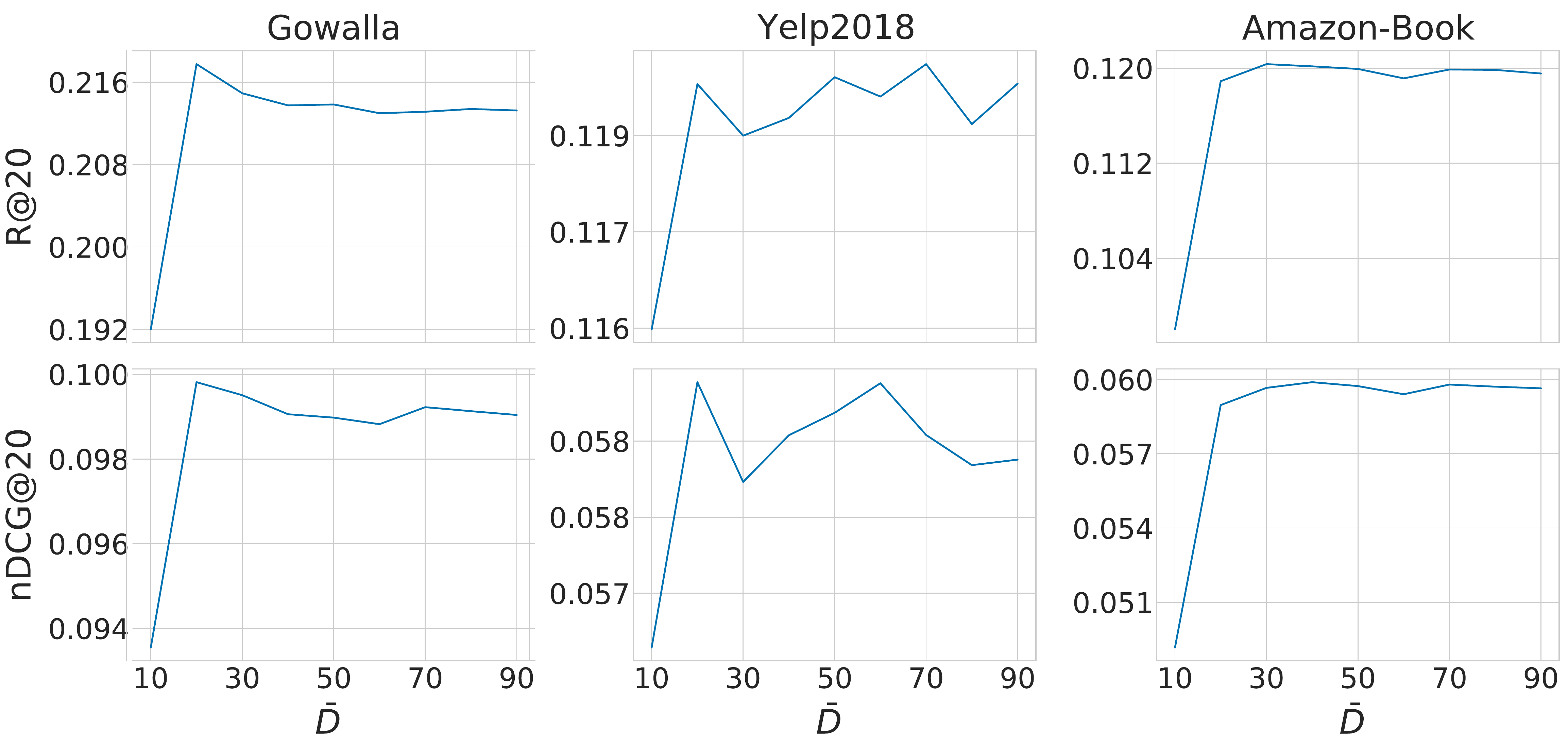}
    \caption{Effect of $\bar{D}$ on R@20 and nDCG@20 on the validation split.}
    \label{fig:sensitivity_C}
\end{figure}

Figures~\ref{fig:sensitivity_lambda}--\ref{fig:sensitivity_C} demonstrate the effects of $\lambda$ and $\bar{D}$, respectively.
in Figure~\ref{fig:sensitivity_lambda}, a large value of $\lambda$ detrimental to R@20 and nDCG@20 for all the datasets, and a small value leads significant deterioration particularly for the Gowalla dataset.
Figure~\ref{fig:sensitivity_C} suggests that the performance is stable when $\bar{D}$ is sufficiently large; these results are consistent to the results in the original work~\cite{kato2020non}.
In conclusion, we recommend tuning both $\lambda$ and $\bar{D}$ for DREGN-CF, although $\lambda$ may be particularly critical for the final performance.

\section{Conclusion}
In this paper, we proposed an LTR approach for personalised recommendation with implicit feedback which reconciles the training efficiency of the pointwise approach and ranking effectiveness of the pairwise approach.
We also demonstrated that the recent non-sampling approach can be considered as PU learning, and our DRE-based approach can be considered as a reasonable alternative of the non-sampling counterpart for top-$K$ recommendation.
We then derived the expected risk based on a weighted variant of the Bregman divergence and proposed a specialised risk for personalised recommendation.
We also provided a scalable implementation including adaptive weighting and user-based sampling strategies, particularly for complex models such as neural networks.
Through empirical analyses on three real-world datasets,
we demonstrated the ranking effectiveness and training efficiency of our method, namely, DREGN-CF.
Our method achieved comparable or substantially higher ranking performance than the method based on the pairwise approach while dramatically reducing the training time.

Particularly for complex models, our DRE-based approach can be a promising alternative of the pairwise counterpart for the following reasons; (1) it can achieve comparable performance to the pairwise counterpart with much shorter training time; and (2) it does not require any extensive sampling strategies.
We expect that our method is beneficial for recommender systems in both real applications and academic researches as it enables rapid development iterations with large-scale data and flexible models with deep learning.
As our method is quite flexible in terms of divergence metric (i.e. $f$), weighting and sampling strategies,
it can be a fundamental method for further advanced methods.

As a future work, we will further explore sophisticated risks that optimise top-$K$ cut-off measures directly;
as our proposed method does not consider the parameter of measures, namely, $K$, the weights in the risk may be improved for a given $K$.

We will publish our program codes of DREGN-CF in the case of acceptance.
The code will be available on GitHub.

\bibliographystyle{ACM-Reference-Format}
\bibliography{head.bbl}

\end{document}